\documentclass[ twocolumn,aps,prd,   
               preprintnumbers,numbers,sort&compress,
               nofootinbib,
                            showpacs,
               colorlinks,
               linkcolor=blue,
               citecolor=blue]{revtex4-1}
   \newcommand{\exclude}[1]{}

\usepackage{graphicx,amsmath,amssymb,bm}
\usepackage{psfrag}
 \usepackage{feynmp}
\usepackage{hyperref}
\usepackage{enumitem}

\newcommand{\diff}[3][{}]{\frac{\mathrm{d}^{#1}{#2}}{\mathrm{d}{#3}^{#1}}}
\newcommand{\vect}[1]{\vec{\mathbf{#1}}}
\providecommand{\d}{}
\renewcommand{\d}{{\rm d}}
\newcommand{\abs}[1]{|#1|}

\begin{document}
\title{Quark Nugget Dark Matter: Comparison with radio observations of nearby galaxies.} 
\author{K. Lawson and A.R. Zhitnitsky}
\affiliation{ Department of Physics \& Astronomy, University of British Columbia, 
Vancouver, B.C. V6T 1Z1, Canada} 

\begin{abstract}
It has been recently 
claimed  that radio 
observations of nearby spiral galaxies essentially rule out a dark matter source for the galactic 
haze\cite{DM-haze-radio}. 
Here we consider the low energy thermal emission from a quark nugget dark 
matter model in the context of microwave emission from the galactic centre and radio observations 
of nearby Milky Way like galaxies. We demonstrate that observed emission levels do not 
strongly constrain this specific dark matter candidate across a broad range of the allowed parameter 
space in drastic contrast with conventional dark matter models based on the WIMP paradigm. 
\end{abstract}

\maketitle

\section{Introduction} 
The galactic microwave `haze' was first detected in the WMAP data \cite{Finkbeiner:2003im} 
and was subsequently confirmed by Planck \cite{Planck-haze}. 
This haze is characterized as diffuse continuum emission, 
centred on the galactic centre and with a harder spectrum than expected for galactic synchrotron 
emission. It is generally believed to be due to the synchrotron emission from the 
injection of a distinct population of high energy particles within the galactic centre which are 
subsequently deflected by the galactic magnetic field. The source of these high energy particles 
has been speculated to be either a recent outburst from the galactic centre or possibly the 
decay or annihilation of dark matter particles into relativistic standard model particles. 

The main motivation for the present work is the claim \cite{DM-haze-radio} that radio 
observations of nearby spiral galaxies essentially rule out a dark matter source for the galactic haze.  
This claim is based on the assumption that if the haze is produced by dark matter annihilation 
or decay, this emission must continue with a similar spectral index down to radio frequencies. 
If this is the case then similar diffuse radio halos should 
exist around other galaxies provided they have a similar distribution of matter and dark matter. 
It is important that all conventional dark matter models based on the weakly interacting 
massive particle (WIMP) paradigm 
do indeed predict a synchrotron spectrum which continues from 
microwave frequencies with $\nu\geq 22$ GHz to radio frequencies with 
$\nu\leq 1$ GHz. Therefore, the assumptions of 
\cite{DM-haze-radio} on the continuity of the spectrum are  well justified for 
WIMP based models. 

Here we study the same question of radio emission from spiral galaxies 
but in a drastically different model, one in which the dark matter  is  represented 
by macroscopically large nuggets of standard model quarks, 
similar to the Witten's strangelets \cite{Witten:1984}, 
see section  \ref{sec:QNDM} for  a short overview of this model. 
In this model  the haze signal is generated by the thermal 
emission from a population of  macroscopically large    nuggets which may constitute the galactic 
dark matter. While this thermal spectrum is similar to the observed haze spectrum across the 
microwave band it falls rapidly at lower frequencies due to many body effects, as we shall argue below. Consequently, the  constraints imposed by radio band emission from 
nearby galaxies is considerably weaker than  
in the case of conventional WIMP type (decaying or annihilating) dark matter models. 
This claim represents the main result of the present work. 

Following a brief review of the properties of the galactic haze (section \ref{sec:haze}) we 
provide an overview of the quark nugget dark matter model   in section \ref{sec:QNDM}. 
With these basics in place we layout the process by which the nuggets may give rise to a 
component of the observed haze emission in section \ref{sec:thermal} and compare the predicted 
spectrum to radio band observations in section \ref{sec:rad_obs}. Our conclusions are presented in 
section \ref{sec:con}.

\section{The galactic haze}
\label{sec:haze}
Initial observations of the haze indicated that it should be considered a unique component of the 
galactic spectrum with a spectral index softer than that of free-free emission and harder than that 
expected for galactic synchrotron. Current estimates 
based on Planck data give a spectral index 
of $\beta_H = -2.55 \pm 0.05$ such that $T_{\nu} \sim \nu^{\beta_H}$ 
\cite{Planck-haze}. In addition to the differing 
spectral index a free-free emission interpretation of the haze is disfavoured by the lack of correlated 
$H\alpha$ emission. Morphologically the haze is 
found to be centred on the galactic centre extending over galactic longitudes $|l| < 15^o$ and 
galactic latitudes $|b| < 35^o$ with an approximately $1/r$ fall off in intensity across that range. 

If the haze is generated by synchrotron emission from a population of relativistic 
particles then there should be a correlated diffuse $\gamma$-ray emission arising from 
the inverse Compton scattering of these particles. Such a component has been detected 
by the Fermi Gamma-Ray Space Telescope \cite{Fermi-haze}. Subsequent observations 
have demonstrated that the $\gamma$-ray component displays relatively sharp edges at high 
latitudes, these features are now referred to as the Fermi bubbles \cite{Fermi-bubbles}. 
Planck also detects a polarized component of the haze which is well correlated with both 
the observed morphology and spectrum of the unpolarized observations 
\cite{Planck-polarization}. This 
combination of features seems to support the idea  that the haze is generated by the injection 
of a population of high energy electrons strongly correlated with the galactic centre. However, 
the source of such a population of relativistic particles remains unknown. A variety of sources 
have been suggested, but they seem to have difficulty describing all aspects of the observed 
emission. In particular the combination of a very sharp edge at large latitudes and strong 
intensity at low latitudes is difficult to reproduce \cite{WMAPhaze_rev}. The sharp edges of the 
bubbles are a noted feature in both the microwave and $\gamma$-ray morphology and strongly 
favour a transient high energy event associated with astrophysical processes in the galactic centre. 
Conversely, the absence of limb darkening at low latitudes favours a process involving the 
continuous injection of the required population of high energy particles. It has been suggested 
that dark matter annihilations or decays may be responsible for injecting these particles. However, 
this interpretation is disfavoured by the sharp bubble edges at large latitudes which do not 
naturally appear in cosmic ray propagation models involving a continuous injection of 
particles. 

The difficulty in reproducing the the morphology of the haze with either 
a transient event in the galactic centre or dark matter emission has also 
lead to the consideration of hybrid models in which only a fraction of the haze 
intensity is provided by dark matter. The analysis of  \cite{Egorov:2015eta} found 
that the  fit to observations is substantially improved if the Fermi bubble correlated 
emission is supported by an additional dark matter contribution at the $\approx 20$\%
level. However, without a well established mechanism for the generation of the 
Fermi bubbles and associated microwave emission any component separation 
remains subject to large uncertainties. 

If the haze is in fact supported by dark matter annihilations or decays then one should 
expect similar emission to be associated with the dark matter halos of nearby galaxies.
Conversely, if the haze is the result of a transient event in the galactic centre there is no 
reason to expect to detect haze like emission from other galaxies. As the haze emission 
generated by relativistic particles injected into the galactic centre will be continuous 
between the microwave and radio bands radio observations of nearby galaxies can be 
used to differentiate between these two models  as argued in  
\cite{DM-haze-radio}. An analysis of nearby spiral galaxies shows 
that they underproduce radio band haze relative to the milky way 
disfavouring the conventional dark matter interpretation of the haze signal  
\cite{DM-haze-radio}. 

With this background in place we will  study the same question  but in a 
drastically different dark matter model 
which  does contribute to the galactic haze but  at the same time is 
not subject to constraints coming from radio band emission, similar to  studies 
of ref.\cite{DM-haze-radio}. 

\section{Quark nugget dark matter}
\label{sec:QNDM}
In this section we will give a brief overview of the quark nugget 
dark matter model. For further details see the original papers 
\cite{Zhitnitsky:2002qa, Oaknin:2003uv, Zhitnitsky:2006vt} 
as well as the recent short review \cite{Lawson:2013bya}. 

The idea that the dark matter  may take the form of composite objects composed of 
standard model quarks in novel phase goes back to stranglet models \cite{Witten:1984}. 
In these models the presence of strange quarks stabilizes quark matter at sufficiently 
high densities, allowing strangelets formed in the early universe to remain stable 
over cosmological timescales. The quark nugget model is conceptually similar, with the 
nuggets being composed of a stable high density colour superconducting phase.
The only new crucial element proposed in \cite{Zhitnitsky:2002qa,Oaknin:2003uv}, 
in comparison with the earlier studies of \cite{Witten:1984} is that the nuggets can be made 
of antimatter as well as matter in this framework.  

The original motivation for this model was unrelated to explaining any 
particular galactic emission source but was related to the seemingly 
unrelated problem of the nature of baryogenesis.
It is generally assumed that the universe 
began in a symmetric state with zero global baryonic charge 
and later (through some baryon number violating process) 
evolved into a state with a net positive baryon number as observed today. 
As an alternative to this scenario we advocate a model in which 
``baryogenesis'' is actually a charge separation process 
in which the global baryon number of the universe remains 
zero. In this model the unobserved antibaryons come to comprise 
the dark matter in form of dense quark (anti) nuggets.  A connection between dark matter and 
baryogenesis is made particularly compelling by the 
similar energy densities of the visible and dark matter 
with $\Omega_{\rm dark} \simeq 5\cdot \Omega_{\rm visible}$. If these processes 
are not fundamentally related the two components could easily 
exist at vastly different scales. 

The observed matter to dark matter ratio corresponds to a scenario   
in which the number  of antinuggets is larger than number of nuggets 
by a factor of $\sim$ 3/2 at the end of the nuggets' formation at the QCD temperature 
$T_{\rm form}\sim \Lambda_{\rm QCD}$ when conventional baryonic visible matter forms. 
This would result in a matter content with baryons, quark nuggets 
and antiquark nuggets in an approximate  ratio 
\begin{equation}
\label{ratio}
B_{\rm visible}: B_{\rm nuggets}: B_{\rm antinuggets}\simeq 1:2:3, 
\end{equation}
with no net baryonic charge.

Unlike conventional dark matter candidates, such as WIMPs the dark-matter/antimatter
nuggets are strongly interacting but macroscopically large.  
They do not contradict any of the many known observational constraints on dark matter 
or antimatter  for three main reasons~\cite{Zhitnitsky:2006vt}:
\begin{itemize} 
\item They carry a huge (anti)baryon charge 
$|B|  \gtrsim 10^{25}$, and so have an extremely tiny number density; 
\item The nuggets have nuclear densities, so their effective interaction
is small $\sigma/M \sim 10^{-10}$ ~cm$^2$/g,  well below the typical astrophysical
and cosmological limits which are on the order of 
$\sigma/M<1$~cm$^2$/g;
\item They have a large binding energy such that baryon charge  in the
nuggets is not available to participate in big bang nucleosynthesis
(\textsc{bbn}) at $T \approx 1$~MeV. 
\end{itemize} 
To reiterate: the weakness of the visible-dark matter interaction is achieved 
in this model due to the small geometrical parameter $\sigma/M \sim B^{-1/3}$ 
rather than due to a weak coupling 
of a new fundamental field with standard model particles. In other words, this small 
effective interaction $\sim \sigma/M \sim B^{-1/3}$ replaces a conventional requirement
of sufficiently weak interactions of the visible matter with  WIMPs. 

\subsection{Nugget properties}
\label{subsec:nug_props}
The nuggets consist of light standard model quarks bound into a colour superconducting state  
in which the quarks form cooper pairs analogous to those found in a traditional 
superconductor. The exact pairing structure, and thus the behaviour of the low energy 
excitations of the quark matter are dependent on the details of the high density QCD 
phase diagram, which remains an open research topic, 
see for example the review \cite{Alford:2007xm}. At large densities the pairing 
of quarks with unique quantum numbers favours the presence of equal numbers of 
u, d and s quarks. However, the relatively large mass of the strange quarks causes  
their relative depletion at lower densities. This results in a net charge for the quark nugget 
(positive in the case of quarks and negative in the case of antiquarks) which is 
compensated for by a layer of leptons known as the `electrosphere' which surrounds 
the nugget. The majority of the observational properties of the nuggets are dictated by the 
properties of the electrosphere, as discussed below, and consequently, are not strongly sensitive to the 
exact details of the quark matter. In particular the low energy thermal emissions that 
are the primary concern of this work are produced in the outer layer of 
the electrosphere, known as the ``Boltzmann" regime, where the positron density has dropped 
sufficiently to become transparent to low energy photons. 

Physically the nuggets will be macroscopically large with a combination of theoretical 
and observational constraints suggesting an average nugget baryon number number 
in the range $10^{25}<B<10^{33}$. Assuming typical nuclear scale densities this translates 
to an average radius in the range $10^{-5}$cm$<R_N<10^{-3}$cm, and to a mass in the 
range from 1g up to thousands of tons. As stated above, the most important physical 
property in terms of scaling the observational consequences of a dark matter model 
is the cross-section to mass ratio. Using standard values for the density of quark matter 
we may estimate,
\begin{equation}
\frac{\sigma}{M} \approx 10^{-10} \frac{{\rm{cm}}^2}{{\rm{g}}} 
\left( \frac{10^{25}}{B} \right)^{1/3}.
\end{equation}

\subsection{Present constraints}
\label{subsec:constraints}
While the observable consequences of this model are on average strongly suppressed  
by the low number density of the quark nuggets the interaction of these objects 
with the visible matter of the galaxy will necessarily produce observable 
effects. Any such consequences will be largest where the densities 
of both visible and dark matter are largest such as the 
core of the galaxy or the early universe. In other words, the nuggets behave as   
conventional cold dark matter in the environment where the visible matter density is 
small, while they become interacting and emitting radiation objects (i.e. effectively 
become visible matter) when in an environment with sufficiently large density.

The features of the nuggets relevant for phenomenology 
are determined by the properties of the electrosphere,  as we have already mentioned. 
The relevant computations can be found   in original refs. 
\cite{Oaknin:2004mn, Zhitnitsky:2006tu,Forbes:2006ba, 
Lawson:2007kp,Forbes:2008uf,Forbes:2009wg,Lawson:2012zu}. 
These properties are in principle, calculable from first principles using only 
the well established and known features  of QCD and QED. As such 
the model contains no tunable fundamental parameters, except for a single 
mean baryon number of the nuggets $\langle B\rangle$.

There are currently a number of both ground based and astrophysical 
observations which impose constraints on  allowed quark nugget dark matter 
parameters. These include the non-detection of a nugget flux by the IceCube 
monopole search \cite{Aartsen:2014awd} which limits the flux of nuggets to 
$\Phi_N < 1$km$^{-2}$ yr$^{-1}$. Similar limits are likely also obtainable 
from the results of the Antarctic Impulse Transient Antenna (\textsc{ANITA}) 
\cite{Gorham:2012an} and it has been suggested that large scale 
cosmic ray detectors may be capable of improving these limits 
\cite{Lawson:2013cr}. While ground based direct searches   
offer the most unambiguous channel for the detection of quark nuggets 
the flux of nuggets is inversely proportional to the nugget mass and 
consequently even the largest available detectors are incapable of 
excluding a nugget flux across their entire potential mass range. 

It has also been suggested that the quark nuggets, through their 
interactions with visible matter, may contribute to astrophysical  
sources of diffuse emission. Analysis of the nugget emission spectrum 
and its consequences in a range of galactic and cosmological environments 
may provide indirect search channels strongly complementary to the 
direct detection searches outlined above. This type 
of analysis has been at least partially carried out for several important 
components of the nugget emission spectrum.  For example, the annihilation of the  
positrons of the electrosphere with incident electrons will 
contribute to the well known galactic 511 keV line \cite{Oaknin:2004mn}.
The positrons farthest from the nugget, with which incoming electrons 
will dominantly annihilate, carry relatively low momenta making the annihilation 
spectrum consistent with the narrow 511 keV line observed \cite{Zhitnitsky:2006tu}. 
As a consequence of the wide range of energy scales involved in the 
electrosphere the 511 keV line will necessarily be accompanied 
by a higher energy ($\sim 10$ MeV) continuum \cite{Lawson:2007kp}. While 
less observational data is available in this range than at either higher or 
lower energies and the astrophysical backgrounds are large 
there is a strong indication of a diffuse galactic excess in the 
MeV range \cite{Strong:2004de}. A detailed analysis of relative annihilation 
rates  suggests that this MeV signal occurs at a level consistent with 
coproduction with the galactic 511 keV line \cite{Forbes:2009wg}. 
At present the uncertainty in the contribution of conventional astrophysical 
processes to the 511 keV line make precise constraints on the 
quark nugget mass difficult to determine. A rough estimate indicates that 
a population of nuggets with $B\sim 10^{24}$ would saturate the observed 
511 keV emission favouring a mean baryon number above this scale. 

The annihilation of galactic protons is a more difficult process to 
study than the relatively simple case of electron-positron annihilations. 
The majority of the released energy is thermalized and emitted as low energy 
radiation to be described in detail below. Even in the case of proton 
annihilations occurring very near the quark surface the energy released 
will rapidly transfer to the many light positrons generating a local hot 
spot on the nugget. This process was analyzed in \cite{Forbes:2006ba} where an x-ray 
band emission signal was predicted. Uncertainty in the background astrophysics 
producing the diffuse x-ray spectrum of galactic centre make  predictions of 
a total nugget contribution across this range difficult to estimate. However, there 
appears to be a hot component to the diffuse x-ray continuum which 
exceeds know astrophysical energy input \cite{Muno:2004bs}. The analysis 
of \cite{Forbes:2006ba} demonstrated the consistency of the proposal that 
the this additional x-ray emission may be coproduced with the galactic 511 keV line. 
As such, while it provides an important consistency check on the quark nugget 
dark matter model it does not significantly improve on the limits obtainable 
from the 511 keV emission strength.  

It has been also suggested recently \cite{Gorham:2015rfa} that  the interactions of 
antinuggets  with normal matter in the Earth and Sun will lead to annihilation and an 
associated neutrino flux. Furthermore, it has been claimed  \cite{Gorham:2015rfa} that 
the antiquark nuggets in the interesting region $10^{25}< B< 10^{33}$  cannot account for 
more than 20$\%$ of the dark matter flux based on constraints for the neutrino flux in 20-50 
MeV range where the sensitivity of the underground neutrino detectors such as SuperK have 
their highest signal-to-noise ratio. However, this claim is based on the assumption that 
the annihilation of visible baryons within an antiquark nugget generates a neutrino spectrum    
similar to the conventional  baryon- antibaryon  annihilation spectrum. In a standard 
baryon- antibaryon annihilation the large number of produced  pions eventually 
decay to muons and consequently to highly energetic neutrinos in the 10-50 MeV energy range. 
The analysis of \cite{Gorham:2015rfa} assumed that the neutrino spectrum 
from the annihilation of an antiquark nugget will fall in this same range. 
However,  this spectrum may be very different for annihilations occurring within 
the colour superconducting nuggets.  
Within most colour superconducting phases the lightest pseudo Goldstone mesons 
(the pions and Kaons) have masses in the 5-20 MeV range \cite{Alford:2007xm} 
considerably lighter than in the  
hadronic confined phase where $m_{\pi}\sim 140$ MeV.  Therefore, these 
light pseudo Goldstone mesons in the 
colour superconducting phase\footnote{We refer to Appendix 2 of 
ref.\cite{Forbes:2006ba}  where it has been explicitly stated that a typical result of the 
annihilation of visible matter with an anti-nugget is the production of very light $m \sim 10$ MeV 
mesons which consequently decay to electrons and neutrinos.}   
will not generally produce highly energetic neutrinos in the 20-50 MeV energy 
range and  thus  are not subject to the SuperK constraints 
employed in \cite{Gorham:2015rfa}. 

\section{Thermal emission}
\label{sec:thermal}
We will now use this basic picture of the quark nuggets and their 
interactions with the surrounding visible matter of the galaxy to extract 
some basic observational consequences of this model in the radio and microwave 
bands relevant to the haze. This will  
involve an analysis of thermal emission from the electrosphere. 

\subsection{Electrosphere}
As discussed in section \ref{subsec:nug_props} thermal emission from the nuggets 
is dominated by the emission of low energy photons from the Boltzmann regime of the 
electrosphere. Consequently our analysis of this emission will require a brief overview 
of the properties of this region, here we essentially follow the results of 
\cite{Forbes:2008uf,Forbes:2009wg}. 

The mean-field approximation for the positron distribution involves solving the Poisson equation
\begin{equation}
\label{eq:Poisson_1}
\nabla^{2}\phi(\vect{r}) = - 4\pi e n(\vect{r})
\end{equation}
where $\phi(\vect{r})$ is the electrostatic potential and $n(\vect{r})$ is the density of 
positrons. As the nuggets are larger than the characteristic scale of the electrosphere 
we are able to work in the one-dimensional approximation 
\begin{equation}
\label{eq:Poisson_z}
\diff[2]{\phi(z)}{z} = - 4\pi e n(z)
\end{equation}
where $z$ is the distance from the quark nugget surface.  We now
introduce the positron chemical potential $\mu_{\text{e}^{+}}(z) =
-e\phi(z)$ which is the potential energy of a charge at position $z$
relative to $z=\infty$ where we take $\mu_{\text{e}^{+}}(\infty) = 0$ as a boundary 
condition. The Poisson equation (equation \ref{eq:Poisson_z}) may then be 
formulated in terms of the chemical potential giving 
\begin{equation}
\label{eq:Poisson}
\frac{\d^2\mu_{\text{e}^{+}}(z)}{\d{z}^2} = 4\pi\alpha\, n[\mu_{\text{e}^{+}}(z)]
\end{equation}
with the additional boundary conditions $\mu_{\text{e}^{+}}(z=0) =
\mu_{0} \sim 10$ MeV as established by beta-equilibrium in the quark matter. 
The full density profile, from the quark surface to vacuum, was computed in 
\cite{Forbes:2009wg}. However, for our analysis it is only necessary to consider 
the low density non relativistic Boltzmann regime where $n\ll (mT)^{3/2}$. In this 
case the positron density is well approximated by
\begin{multline}
\label{eq:Boltzman_n}
n[\tilde{\mu}] \approx 2\int \frac{\d^{3}{p}}{(2\pi)^3}\;
e^{\frac{[\tilde{\mu} - p^2/(2m)]}{T}} 
= \sqrt{2}\left(\frac{mT}{\pi}\right)^{3/2}e^{\frac{\tilde{\mu}}{T}}.
\end{multline}
The effective chemical potential $\tilde{\mu} = \mu_{\text{e}^{+}} - m$ is related to the 
vacuum chemical potential $\mu$ by subtracting the mass.  We note that the right
boundary condition must now be changed to $n(z=\infty) = 0$ because
$\tilde{\mu}$ does not tend to zero under these approximations.  The
left boundary condition must be determined by matching the density at
some point to the full relativistic solution that integrates to the
quark-matter core.
The differential equation~(\ref{eq:Poisson}) has  the peculiar solution
\begin{equation}
\label{eq:n_z}
n(z) = \frac{T}{2\pi\alpha}\frac{1}{(z+\bar{z})^2},
\end{equation}
where $\bar{z}$ is an integration constant fixed by matching to a
full solution. A proper computation of $\bar{z}$ would require 
tracking the density through many orders of magnitude from the
ultrarelativistic down to the nonrelativistic regime.  These computations 
as we already mentioned, have been carried out in  \cite{Forbes:2009wg}. 
However, a simple approximation will suffice for present purposes. 
We take $z=0$ to define the onset of the Boltzmann regime: 
\begin{eqnarray}
\label{eq:z=0}
n(z=0) &=& \frac{T}{2\pi\alpha\bar{z}^2} = (mT)^{3/2}, \nonumber\\  
\frac{1}{\bar{z}} &\simeq& \sqrt{2\pi\alpha}\cdot m \cdot\sqrt[4]{\frac{T}{m}}. 
\end{eqnarray}
Numerically, $\bar{z}\sim 0.5\cdot 10^{-8}$~cm while the density
$n\sim 0.3 \cdot 10^{23} $~cm$^{-3}$ for $T\simeq 1$eV.
A comparison with the exact numerical results  of \cite{Forbes:2009wg}
support  our approximate treatment of the problem in terms of  parameters represented by  
equations (\ref{eq:n_z}) and (\ref{eq:z=0}). In this formulation the region 
described by $z<0$ corresponds to the high density regime where the Boltzmann 
approximation breaks down and which is opaque to thermal photons \cite{Forbes:2009wg}. 

\subsection{Thermal spectrum and  \textsc{LPM} suppression}
\label{subsec:spec} 
In order to determine the thermal spectrum we begin estimating the 
emissivity of the positrons of the Boltzmann regime, this calculation follows 
the results of \cite{Forbes:2008uf} but will provide a more careful treatment of 
the low energy behaviour than was required in that analysis. 

The starting point is  the following expression for the cross
section for two positrons emitting a soft photon with $\omega\ll p^2/(2m)$, see 
\cite{Forbes:2008uf} for the detail discussions on validity  of this classical formula, 
\begin{equation}
\label{eq:sigma}
\d\sigma_{\omega}=\frac{4}{15}\alpha \left(\frac{\alpha}{m}\right)^2
\cdot\left(17+12\ln\frac{p^2}{m\omega}\right)\frac{\d\omega}{\omega}.
\end{equation}
The emissivity $Q=\d{E}/\d{t}/\d{V}$---defined as the total energy
emitted per unit volume, per unit time---and the spectral properties
can be calculated from
\begin{multline}
\label{eq:spectrum}
\frac{\d{Q}}{\d{\omega}}(\omega,z) = n_1(z, T)n_2(z, T)  \omega
\left\langle v_{12}\frac{d\sigma_{\omega}}{d\omega}\right\rangle\\
=\frac{4\alpha}{15}\left(\frac{\alpha}{m}\right)^2n^2(z,T)
\left\langle v_{12}\left(17+12\ln\frac{p_{12}^2}{m\omega}\right)\right\rangle
\end{multline}
where $n(z,T)$ is the local density at distance $z$ from the nugget's
surface, and $v_{12} = \abs{\vec{v}_1-\vec{v}_2}$ is the relative
velocity.  The velocity and momentum $p_{12}$ need to be thermally
averaged.  Assuming the Boltzman distribution (\ref{eq:Boltzman_n}) the corresponding 
computations lead to the following expression  \cite{Forbes:2008uf}:
\begin{multline}
\left\langle v_{12}\left(17+12\ln\frac{mv_{12}^2}{\omega}\right) \right\rangle =\\
= 2\sqrt{\frac{2T}{m\pi}} \left(1+\frac{\omega}{T}\right)
e^{-\omega/T} h\left(\frac{\omega}{T}\right)
\end{multline}
where the function  $h(x)$ for all $x$ can be approximated as follows
\begin{equation}
h(x) = \begin{cases}
17-12\ln(x/2) & x<1,\\
17+12\ln(2) & x\geq1.
\end{cases}
\end{equation} 
Therefore , the emissivity $Q$ assumes the form
\begin{eqnarray}
\label{eq:Q}
\frac{\d{Q}}{\d{\omega}}(\omega,z) &=&\frac{8\alpha}{15}\left(\frac{\alpha}{m}\right)^2n^2(z,T) \\
&\times&\sqrt{\frac{2T}{m\pi}} \left(1+\frac{\omega}{T}\right) e^{-\omega/T}
h\left(\frac{\omega}{T}\right).
\end{eqnarray}

To translate this volume emissivity into a spectral surface emissivity we 
integrate over the positron density distribution given in expression (\ref{eq:n_z}). 
The resulting surface emissivity ($F \equiv \int\d{z}\;Q(z)$) is defined as the energy  
emitted per unit time, per unit surface area at a given frequency. The result of 
integrating expression (\ref{eq:spectrum}) is, 
\begin{multline}
\label{eq:P}
\frac{\d{F}}{\d{\omega}}(\omega) = 
\frac{\d{E}}{\d{t}\;\d{A}\;\d{\omega}}
\simeq
\frac{1}{2}\int^{\infty}_{0}\!\!\!\!\!\d{z}\;
\frac{\d{Q}}{\d{\omega}}(\omega, z)
\sim \\
\sim
\frac{4}{45}
\frac{T^3\alpha^{5/2}}{\pi}\sqrt[4]{\frac{T}{m}}
\left(1+\frac{\omega}{T}\right)e^{-\omega/T}h\left(\frac{\omega}{T}\right),
\end{multline}
where factor $1/2$ accounts for the fact that only the photons emitted
away from the core can actually leave the system. Integrating over $\omega$ 
contributes a factor of $T\int\d{x}\;(1+x)\exp(-x)h(x)\approx 60\,T$, 
giving the total surface emissivity:
\begin{equation}
\label{eq:P_t}
F_{\text{tot}} = \frac{\d{E}}{\d{t}\;\d{A}} = 
\int^{\infty}_0\!\!\!\!\!\d{\omega}\;
\frac{\d{F}}{\d{\omega}}(\omega) 
\sim \frac{16}{3}
\frac{T^4\alpha^{5/2}}{\pi}\sqrt[4]{\frac{T}{m}}. \\
\end{equation}
From equation (\ref{eq:P}) it is clear that emission from the nuggets 
will be peaked at frequencies near $\hbar \omega \sim T$ and displays 
a weak (logarithmic) dependence on frequency when $\hbar \omega << T$. 

This derivation is identical to that provided in \cite{Forbes:2008uf} which 
analyzed thermal emission from nuggets within the galactic centre. That analysis 
focussed on a possible contribution to the galactic spectrum from the nuggets in the 
microwave range. As we now want to consider radio band emission it is necessary 
to treat the low energy tail of the spectrum more carefully. 

One may ask how microwave radiation   may be emitted from the nuggets
when the wavelength $\lambda$ is much larger than the size of the
nugget $\lambda\gg R$.  In general this is not a problem---consider the
well-known astrophysical emission of the $\lambda= 21$~cm line from
hydrogen with a size $a\simeq 10^{-8}$~cm. This example shows that
important part of the question is not the size of the system but rather, the coherence time.
 The coherence time $\tau$ of the positrons which must be
compared with the formation time $\sim\omega^{-1}$ of the photons.  If
the coherence time is too short, then multiple scatterings will
disrupt the formation of the photons.  This suppression is a case of
the so-called Landau-Pomeranchuk-Migdal (\textsc{lpm}) effect \cite{LPM}, see also
recent application  of the \textsc{lpm} effect in similar context of  quark dense stars \cite{Jaikumar:2004rp}.  

To estimate the coherence time $\tau$ for our case, consider the cross-section
$\sigma_{ee}$ of the positron-positron interaction.  This scales as
$\sigma_{ee}\sim \alpha^2/q^2$ where $q\sim b^{-1}$ is the typical
momentum transfer, and may be expressed in terms of the impact
parameter $b\sim n^{-1/3}$, which is estimated in terms of average
interparticle spacing where $n$ is the local positron density.
The mean-free-path $l$ is thus $l^{-1}\sim \sigma_{ee}n\sim \alpha^2n^{1/3}$.
Therefore, the typical time between collisions (which is the
same as coherence time) is $\tau\sim l/v$ where $v\sim
\sqrt{T/m}$ is the typical positron velocity.

Collecting all factors together and using (\ref{eq:n_z}) for the density profile we arrive at the estimate
\begin{equation}
\label{eq:coherence}
\omega\tau \sim \frac{\omega}{\alpha^2n^{1/3}}\sqrt{\frac{m}{T}}
\sim \frac{\omega}{\alpha^2T}\left( 1+\frac{z}{\bar{z}}\right)^{\frac{2}{3}}
\geq 1.
\end{equation}
One can check  that this condition is  satisfied for $\omega\geq
10^{-4}$~eV and $T\leq 1$~eV even for $z=0$.  Thus, we were  marginally justified
in omitting \textsc{lpm} effect in our estimates (\ref{eq:P}) in the low-density
regime (\ref{eq:n_z}) for $\omega\geq
10^{-4}$~eV, which corresponds to the longest wave length with $\nu \geq 22$ GHz
in WMAP haze studies. However, from the same estimate it is clear
that this suppression becomes important for  smaller frequencies $\omega \ll 10^{-4}$~eV.   

We want explicitly  take into account the corresponding suppression for 
radio waves with $\nu \ll 20$ GHz. One can implement this suppression 
into our formula (\ref{eq:P}) as follows. First, consider  the minimal frequency 
when condition (\ref{eq:coherence}) is marginally satisfied for  $z\geq z_{min}$, i.e. 
\begin{equation}
\label{eq:omega_0}
\frac{\omega}{T} 
= \alpha^2\left(\frac{\bar{z}}{z_{min}+\bar{z}}\right)^{\frac{2}{3}}, ~~ \omega_0=\alpha^2T.
\end{equation}
For sufficiently large frequencies  $\omega\geq \omega_0  $ the \textsc{lpm} 
effect is not operational anywhere in electrosphere  even for $z=0$.  
In this case one can integrate over entire region $\int_0^{\infty}dz$ of the 
electrosphere.   This  is precisely the procedure leading to eq.    (\ref{eq:P}).

However, for radio   frequencies $\omega\leq\omega_0$
the  \textsc{lpm} effect is operational, at least in some region of $z$. This effect  
strongly suppresses the emission of the low energy photons from that region.  We 
want to account for this suppression using the following technical trick. We 
separate the integral entering  (\ref{eq:P}) into two regions, the high density region, 
and the low density region correspondingly: 
\begin{eqnarray}
\label{eq:integral_LPM}
\int^{\infty}_{0}\!\!\!\!\!\d{z}\;  \frac{\d{Q}}{\d{\omega}}
=  \int^{z_{min}}_{0}\!\!\!\!\! \d{z}\;  \frac{\d{Q}}{\d{\omega}}+\int^{\infty}_{z_{min}}\!\!\!\!\!\d{z}\;  
\frac{\d{Q}}{\d{\omega}}.
\end{eqnarray}
Within the high density region $z\leq z_{min}(\omega)$ the coherence condition 
(\ref{eq:coherence}) is not satisfied and the production of low energy photons 
is strongly \textsc{lpm} suppressed. Conversely, for $z\geq z_{min}$ the coherence 
condition is satisfied and photon production proceeds essentially as in vacuum. 
\begin{figure}
\includegraphics[width=\linewidth]{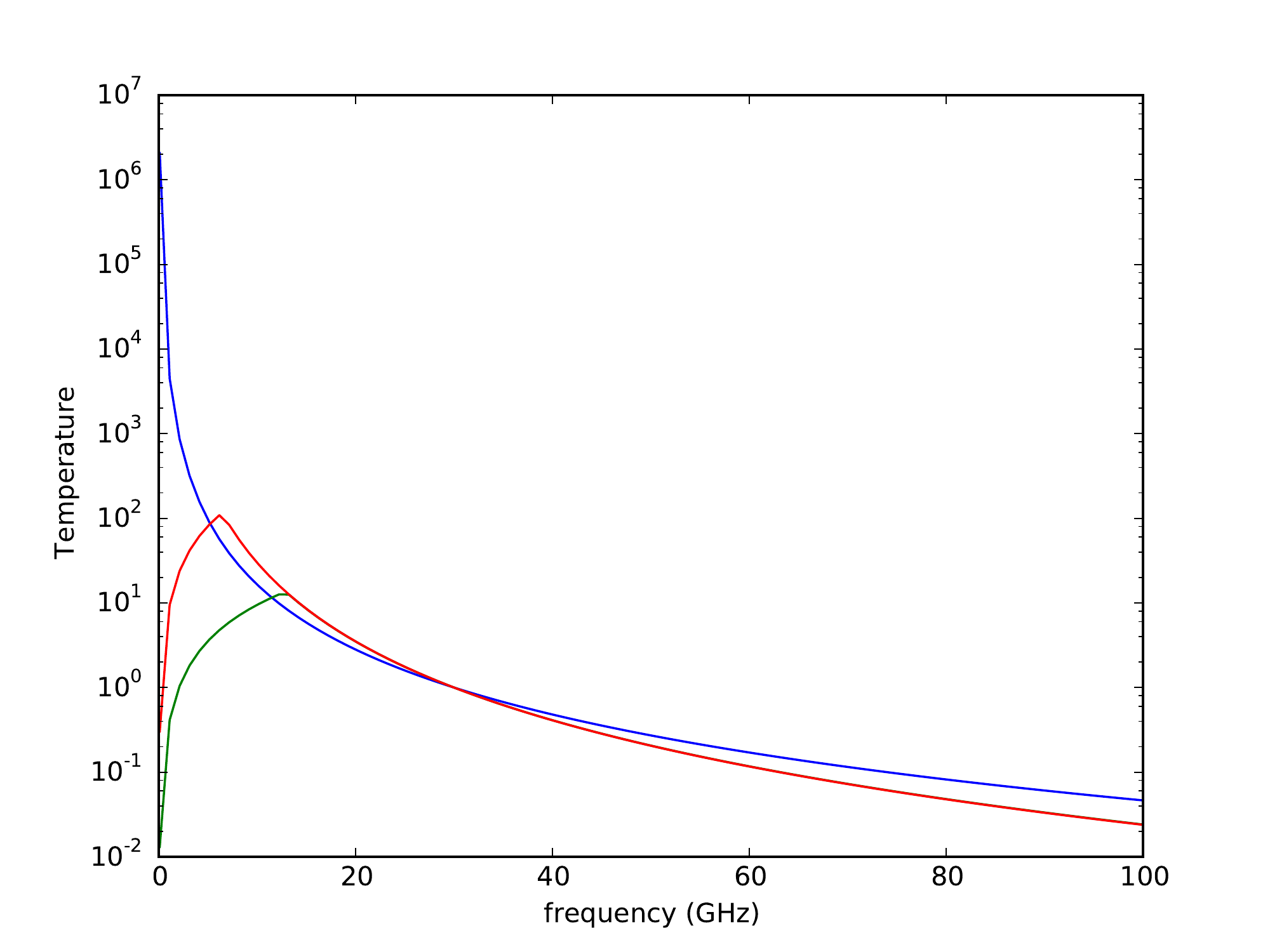}
\caption{The spectrum of quark nuggets across the tens of 
GHz range for nugget temperatures of $T_N = 0.5$eV (red) and 
$T_N = 1$eV (green). Also shown is the power law spectrum reported 
by Planck with $T \sim \nu^{-2.55}$ continued to the low energy region with 
the same spectral index. All spectra are normalized at 
$\nu = 30$GHz, as such that the total magnitude is arbitrary.}
\label{fig:spec}
\end{figure}
For our estimate it is sufficient to disregard the emission of the low energy photons 
from  the high density region and focus on emission from $z\geq z_{min}$. 
In other words, the region of integration in our computation of the spectral 
surface emissivity  (\ref{eq:P}) becomes frequency dependent, 
\begin{eqnarray}
\label{eq:cutoff} 
\frac{\d{F}}{\d{\omega}}(\omega) \simeq 
\frac{1}{2}\int^{\infty}_{z_{min}} \d{z}
\frac{\d{Q}}{\d{\omega}}(\omega, z) ~~{\rm for } ~~\omega\leq \omega_{0}, 	 
\end{eqnarray}
where $z_{min}$ depends on $\omega$ according to (\ref{eq:omega_0}). 
Formula (\ref{eq:cutoff}) is identical to our original formulation (expression \ref{eq:P}) 
for high frequency photons while at low frequencies an ever smaller fraction 
of the electrosphere contributes to the surface emissivity. The $z$ integration in 
expression (\ref{eq:cutoff}) can be easily  computed as the density profile has a 
simple analytical expression as a function of $z$ in the Boltzman regime (\ref{eq:n_z}). 
The resulting suppression factor  $\Delta(\omega)\leq 1$ for $\omega\leq \omega_0$ 
is convenient to represent as follows:
\begin{eqnarray}
\label{eq:suppression} 
\Delta (\omega)&=& \left(\frac{\bar{z}}{z_{min}+\bar{z}}\right)^{3} \simeq
\left(\frac{\omega}{\alpha^2T} \right)^{\frac{9}{2}}\nonumber \\
&\simeq&
\left(\frac{\omega}{\omega_0}\right)^{\frac{9}{2}}\left(\frac{T_0}{T}\right)^{\frac{9}{2}}, 
\end{eqnarray}
where $\omega_0\simeq 10^{-4} $eV for $T_0\simeq 1$ eV. 
In deriving (\ref{eq:suppression}) we used the fact that  
the integral entering  (\ref{eq:cutoff}) with density profile (\ref{eq:n_z}) leads to the 
cubic dependence on cutoff as shown in (\ref{eq:suppression}). The corresponding 
density cutoff is further expressed  in terms of frequency of emission $\omega$  
according to (\ref{eq:omega_0}). 

The combination of the spectrum given in equation (\ref{eq:P}) and the 
\textsc{lpm} suppression factor of equation (\ref{eq:suppression}) allow 
us to describe the thermal spectrum of the nuggets from the eV scale down 
to radio frequencies. This spectrum is plotted in Fig.\ref{fig:spec} showing its 
similarity to the reported haze spectrum in the tens of GHz range as well as the 
low energy cutoff. 

One should remark here that our treatment of the low frequency part of the spectrum
at $\omega\leq \omega_0$ is equivalent to very sharp ``removal" of the corresponding 
emission from the high density region with $z\leq z_{min}(\omega)$. In Fig.\ref{fig:spec}
this corresponds to almost ``cusp" like behaviour of the spectrum. In reality the  
\textsc{lpm} suppression becomes operational  in the extended region of 
$z\simeq z_{min}$ with the typical width $\Delta z\simeq \bar{z}$ according to 
eq. (\ref{eq:suppression}). The cusp in Fig.\ref{fig:spec} will be smoothed out by this modification. 
However, the basic qualitative behaviour is unaffected by this smoothing and remains 
the same as plotted in Fig.\ref{fig:spec}. A precise treatment of this transition region where  
the \textsc{lpm} effect becomes operational  is a technically challenging problem. 
Fortunately, for our purposes we do not need the precise form of this transition region. 
Therefore, we will use our rough  estimates in their present form for the following analysis. 

To reiterate: Our procedure employed above obviously introduces some numerical uncertainty 
of order unity in the suppression factor (\ref{eq:suppression}).
However, this expression  obviously shows that the radio wave emission is strongly 
suppressed by this mechanism, while emission at CMB frequencies with 
$\omega\geq \omega_0$ remain essentially untouched by this suppression.  

\subsection{Other potential correction  factors in radio emission bands}
\label{subsec:others}
The previous subsection  analyzed the key factor, the \textsc{lpm} suppression 
which influences the radio emission from dark matter nuggets, which is the main 
subject of the present studies. We now want to consider some other sources which 
may also affect  the low energy  radio emission.

1.  The mean-field approximation which we explored in deriving expression 
(\ref{eq:n_z}) is not valid for extremely large $z$, where exponential
rather than power-law decay is expected.  We could
accommodate the corresponding feature by introducing a cutoff at some 
sufficiently large $z=z_{\text{max}}$ on the order the radius  of the nugget 
$R\sim 10^{-5}$cm.  The result, however, is not sensitive to this cutoff, so we use 
$z_{\text{max}}=\infty$  in our formula (\ref{eq:P}). This cutoff at very large 
$z=z_{\text{max}}$ does not affect our expression for 
the suppression factor (\ref{eq:suppression}) because at large $z_{\text{max}}$ 
the positron density is already small enough to contribute little to the overall emissivity.

2. Our calculations have assumed that we are working in infinite
matter. However,  the nuggets have a finite extent on the order of 
$R \geq10^{-5}$~cm.  
In principle, finite-size effects may change the positron scattering cross-section
(\ref{eq:sigma}), and therefore, our estimation of the emissivity
(\ref{eq:spectrum}).  The cross-section (\ref{eq:sigma}) was derived
using a continuum of plane-wave states, whereas to account for the
finite-size effects, one should use the basis of states bound to the
quark core.  To estimate the size of the corrections, one can imagine
confining the positrons to a box of finite extent $R$.
The electromagnetic field may still be quantized as in free-space with
states of arbitrarily large size because the photons are not bound to
the core, and are not in thermodynamic equilibrium with the positrons.
Their mean-free-path is much larger than $R$, so the low-energy
photons produced by the mechanism described above will simply leave
the system before they have a chance to interact with other positrons.

Therefore, it is only the positron states that must be considered over
a finite-size basis, which will modify the corresponding Green's
function used in the calculation of the cross-section
(\ref{eq:sigma}).  These modifications occur for momenta of the scale
$\delta p\sim \frac{n\hbar}{R}$ with $n$ being an integer number describing 
the typical excitation level.  If $R \geq 10^{-5}$~cm, then this corresponds
to shifts in the energies of $\delta E \sim (\delta p)^2/2m \sim
10^{-6}$~eV~$\ll 10^{-4}$~eV, which is much smaller than the
transitions responsible for the emission at microwave frequencies. 
One could naively think that this energy shift could affect 
emission at radio frequencies $\omega \sim 10^{-6}$~eV, which is the main 
subject of the present work. However, this is not the case
because the typical positron energy scale is determined by the nugget 
temperature $T_n\sim 1$eV corresponding to very large excitation 
numbers  $n\gg 1$ for the positrons responsible for emission.
Thus, we conclude that finite-size effects do not drastically change
the positron Green's function in the region of interests.  In
other-words, the expression for the cross-section
(\ref{eq:sigma})---derived using the standard (infinite volume)
Green's functions---remains valid for our estimation of the emission
and spectrum down to radio frequencies when $T_n\sim 1$ eV, and we may use our  
original expressions for the emissivity (\ref{eq:Q}) and suppression (\ref{eq:integral_LPM}), 
(\ref{eq:suppression}).  We also note
that finite-size effects do not change our estimates for the density
(\ref{eq:Boltzman_n}) because the finite-size effects
$\delta{E} \ll T$ are much smaller than the typical energetic scale $T_n\sim $~eV 
of the problem.  Thus, our expression (\ref{eq:sigma}), and therefore (\ref{eq:P})
remains valid even for radio frequency photons which is the main subject of the present studies. 
 
3.  Another   factor which may  potentially affect the  low energy emission from the nuggets is the 
generation of the plasma frequency  $\omega_p^2=\frac{4\pi\alpha n}{m}$ 
in the electrosphere. The plasma frequency can be thought as an effective mass
for the photon: only photons with energy larger than this mass can  
propagate within the system and eventually escape 
the nugget.  Photons with $\omega < \omega_{p}$ 
are ``off-shell'' or ``virtual'': these can only propagate for a short
period of time/distance $\sim \omega_{p}^{-1}$ before they decay (are absorbed).
This effect, similar to the \textsc{lpm} effect, also suppress the low energy emission. 
However, the physics  of generating  the plasma frequency  are different  from  
those involved in the \textsc{lpm} effect discussed in the 
previous section \ref{subsec:spec}.  The observable manifestations 
of this phenomenon are also different from the \textsc{lpm} effect --the low energy photons, 
even if they are produced,  can not propagate in an environment with non vanishing 
$\omega_{p}$. This should be contrasted with \textsc{lpm} effect 
in which low energy photons cannot be even formed.  

One can estimate that the plasma frequency $\omega_{p}$ is in the few eV range
for densities (\ref{eq:n_z}) at $z= 0$  and even smaller for large $z>0$.  Given
our previous discussion, one might ask: How can low-energy photons
$\omega < \omega_{p}$ which are the subject of the present work, still be emitted?  
The reason is that, although
these photons would be reabsorbed in infinite matter, this
reabsorption happens on a length scale of $\omega_p^{-1}$.  At the
typical densities in the Boltzmann regime, $\omega_{p}^{-1}\sim0.3\cdot 10^{-5}$~cm 
is much larger than  $\bar{z}\sim 10^{-8}$ cm where such high density
is supported by the nugget's structure (\ref{eq:z=0}).  Therefore,  many of these 
photons will have left the nugget before being reabsorbed.    
Therefore, this effect is important in the deep dense regions of the nuggets.
It would be also important if our system would be infinitely large.  
However, the generating of the plasma frequency  $\omega_{p}$  does 
not affect  our expression for the emissivity (\ref{eq:Q}) and corresponding  estimates 
(\ref{eq:integral_LPM}), (\ref{eq:suppression})  for  finite size  nuggets the in radio bands,  
which is the subject of our present studies. 
 
To conclude this section: with the estimates just presented we are now 
in position  to consider the potential for radio band observations to search 
for the presence of quark nugget populations within our own or nearby galaxies.

\section{Radio band intensity calculations}
\label{sec:rad_obs} 
The emission spectrum of a quark nugget within a given environment is 
determined by its temperature. In the case of an antiquark nugget the primary 
heating mechanism is the annihilation of visible matter within the nugget\footnote{Nuggets 
composed of quarks rather than antiquarks will experience purely collisional heating 
and will be at a much lower temperature. Consequently we may safely neglect their 
impact on the galactic spectrum.}. Within the galactic interstellar medium (\textsc{ism}) 
the flux of matter onto the nuggets is simply the product of the local visible matter 
density and the mean velocity. The total heating rate of the nugget is then given by, 
\begin{equation}
\label{eq:E_in}
\frac{\d{E}}{\d{t}} = \rho_{\rm vis} v f_T \sigma_N
\end{equation}
where $f_T$ is the fraction of colliding mass which annihilates and thermalizes 
within the nugget and $\sigma_N$ is the nugget cross-section. Equating this 
heating rate with the rate of thermal emission from equation (\ref{eq:P_t}) gives 
the nuggets' radiating temperature in a given environment:
\begin{equation}
\label{eq:T_N}
T_N = 0.5~ {\rm{eV}} 
\left[ \frac{\rho_{\rm vis}}{10{\rm{GeV}}/{\rm{cm}}^3} ~\frac{v}{200 {\rm{km}}/{\rm{s}}} ~
f_T \right]^{4/17}.
\end{equation}
This temperature fixes the emission spectrum of an individual nugget. Using eqs.(\ref{eq:P}) and (\ref{eq:P_t}) the corresponding spectrum can be written in the following form,
\begin{equation}
\frac{\d{E}}{\d{t}~\d{\omega}} = 
\frac{\rho_{\rm vis} v \sigma_N f_T}{60 T} 
\left( 1 + \frac{\hbar \omega}{T} \right) e^{-\hbar \omega/T} 
h\left( \frac{\hbar \omega}{T} \right) .
\end{equation}
The volume emissivity of the \textsc{ism} due to the presence of 
quark nugget dark matter is then given by scaling the individual nugget 
spectrum by the number density of nuggets,
\begin{eqnarray}
\label{eq:nug_emiss}
\epsilon_N &\equiv& \frac{\d{E}}{\d{\omega}~\d{t}~\d{V}} 
= \frac{\rho_{DM}}{M_N} \frac{\d{E}}{\d{\omega}~\d{t}} \nonumber \\ 
&=& \frac{\rho_{\rm vis} v \sigma_N \rho_{DM} f_T}{90 M_N T} 
\left( 1 + \frac{\hbar \omega}{T} \right) e^{-\hbar \omega/T} 
h\left( \frac{\hbar \omega}{T} \right)
\end{eqnarray}
where $M_N$ is the average quark nugget mass and we have included a factor 
of $2/3$ to account for the fact that only the antiquark nugget component of the 
dark matter will contribute to the radio band spectrum. As established in the low 
frequency treatment of section \ref{subsec:spec} expression (\ref{eq:nug_emiss}) 
must be multiplied by the suppression factor (\ref{eq:suppression}) for frequencies 
below $\omega \sim \alpha^2T$. 

Note that the physical properties of the nuggets entering into expression (\ref{eq:nug_emiss}) 
are carried by the cross section to mass ratio $\sigma_N/M_N$. There is also a dependence on 
the thermalization coefficient $f_T$ both as an overall scaling factor and through the dependence 
of emissivity on the radiating temperature (from equation \ref{eq:T_N}) however the value of 
$f_T$ is expected to fall in the range $1>f_T>1/2$ so this factor contributes only 
marginally when compared to the much larger allowed range of $\sigma_N/M_N$. Note 
that $\sigma_N \sim B^{2/3}$ while $M_N \sim B$ so that the cross section to mass ratio 
scales with the nugget baryon number as $B^{-1/3}$. As  already mentioned in the Introduction 
this small geometrical factor replaces the weakness of the visible-dark matter interaction 
in conventional WIMP paradigm. 

\subsection{Matter distributions}
\label{subsec:matter_profiles}
The emissivity given in equation (\ref{eq:nug_emiss}) allows us to determine 
the thermal emission from a population of quark nuggets provided we know the 
distribution of matter and dark matter. We will adopt the standard  
Navarro-Frenk-White (\textsc{nfw}) profile, 
\begin{equation}
\label{eq:NFW}
\rho_{NFW}(r) = \rho_s \left( \frac{r_s}{r} \right) 
\left( 1 + \frac{r}{r_s} \right)^{-2}
\end{equation}
so that the dark matter profile of a given galaxy may be described by the 
scale radius ($r_s$) and the characteristic density ($\rho_s$). For example the dark matter 
halo of the Milky Way is generally taken to have $r_s \approx 22$ kpc and 
$\rho_s \approx 0.5$GeV/cm$^3$. The visible matter distribution 
is generally more complicated and, for present purposes, we will attempt to capture only 
its basic properties. Of primary importance in the context of dark matter interactions 
is the central, spherically symmetric, galactic 
bulge. We will model the bulge with a simple exponential, 
\begin{equation}
\label{eq:bulge}
\rho_B(r) = \rho_0 e^{-r/r_0}
\end{equation}
with central density $\rho_0$ and scale length $r_0$. For a Milky Way like galaxy we 
expect $r_0 \approx 3$kpc and $\rho_0 \approx 100$GeV/cm$^3$. 
Additionaly we will include a disk component for the visible matter, 
\begin{equation}
\label{eq:disk}
\rho_d(h) = \rho_d e^{-h/H_0}
\end{equation}
where $h$ is the height above the galactic plane, $\rho_d$ is the in plane density 
and $H_0$ is the disk scale height. For a Milky Way like spiral we may estimate the 
central disk density as $\rho_d \approx 1$GeV/cm$^3$ and a disk scale height of 
$H_0 \approx 0.5$kpc. The disk distribution will be cut off at a maximum 
distance $d_{max}$ from the galactic centre. 

The final property of the galactic matter distribution we need is the average velocity.
While some galactic matter has been significantly accelerated the majority carries a 
velocity on the order of the galactic rotation speed. As such we will consider the average 
velocity of the matter populations to be on the order of $v \sim 200$km/s. 

\subsection{Milky Way}
\label{subsec:MW}
\begin{figure}
\includegraphics[width=\linewidth]{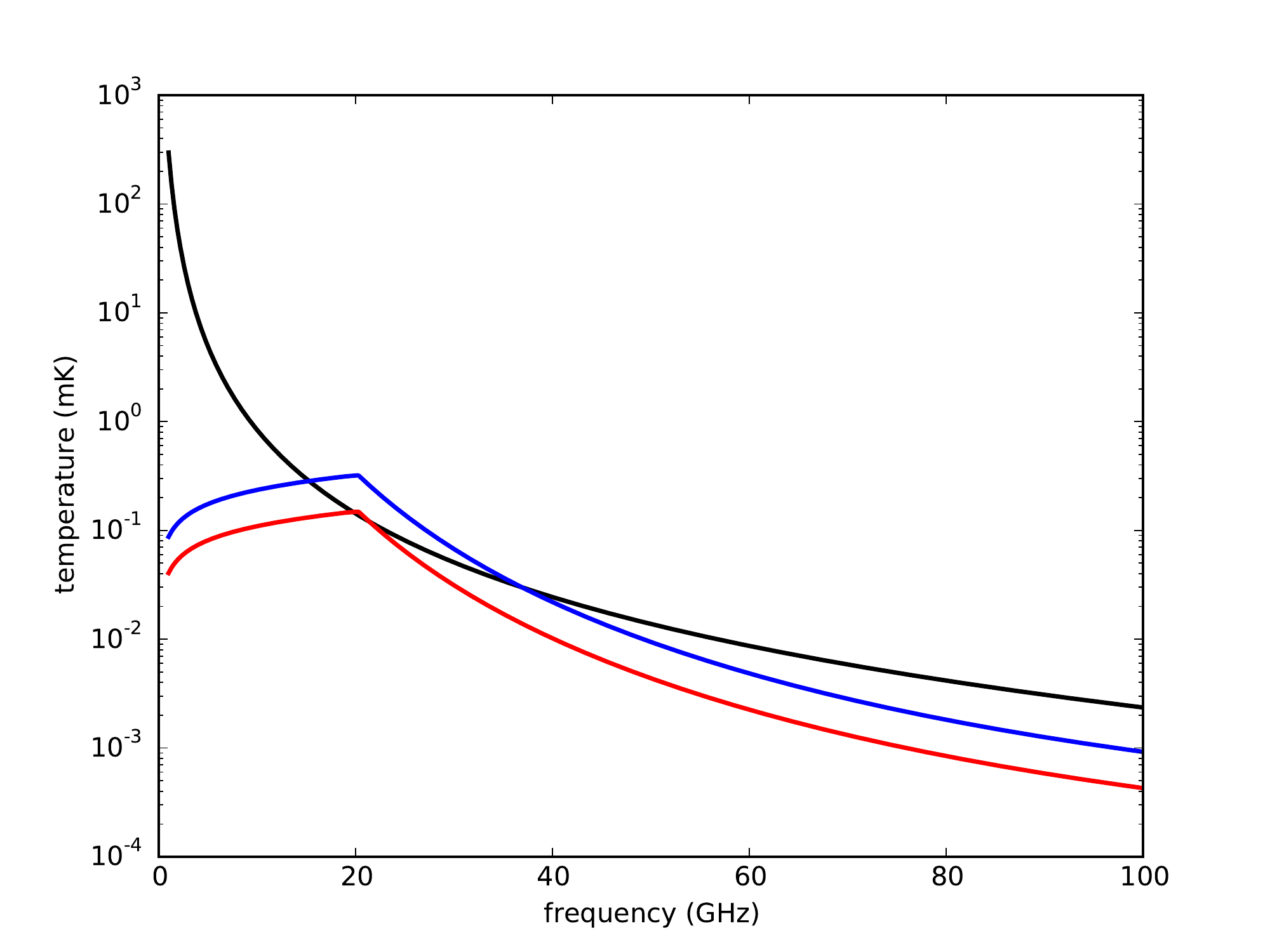}
\caption{The spectrum of quark nuggets across the tens of 
GHz frequencies observed by \textsc{WMAP} and Planck. The haze spectrum 
as reported by Planck is plotted in black, continued to the low energy region 
with the same spectral index $T\sim \nu^{-2.55}$.
The spectra for a quark nugget population 
with $B\sim10^{25}$ is plotted in blue and that of a population with $B\sim10^{26}$
is shown in red. The nugget contribution that from the $B\sim10^{25}$ population 
would saturate the haze emission from the galactic centre and, as such, any 
mean baryon number below this value is effectively 
ruled out by the current Planck data. See text for more specific discussion of limits.}
\label{fig:gc_spec}
\end{figure}

The flux received from the quark nugget population within our galaxy may be determined 
by the integral of the emissivity given in expression (\ref{eq:nug_emiss}) along a given line 
of sight. We are particularly interested in the intensity received from the direction of the 
galactic centre where both the dark and visible matter distributions are strongly 
peaked. For simplicity we here consider ignore the visible matter in the disk and 
focus on the bulge component which strongly dominates along lines of sight through 
the galactic centre. This introduces a rotational symmetry and somewhat simplifies 
the integration procedure. In this case the flux received from a line of sight through 
the galactic centre is given by 
\begin{equation} 
\label{eq:gc_los}
\Phi = \int \frac{\d{V}}{4\pi r_\odot^2} \epsilon(r_g), 
\end{equation}
where $r_\odot$ is the distance from earth and $r_g$ is total distance 
from the galactic centre. Exploiting the rotational symmetry of the problem 
this may be simplified to give 
\begin{equation}
\label{eq:gc_los_int}
\Phi = \int_0^{\infty}\d{r} \int_0^{h_{max}} \frac{h\d{h}}{h^2 + r^2} \epsilon(r_g)
\end{equation}
where $r$ is radial distance from earth along the galactic plane and $h$ is height 
above the plane. Thus $r_g \equiv \sqrt{ (R_\odot-r)^2 + h^2}$ where $R_\odot$ is the 
earth's distance from the galactic centre. The maximum height ($h_{\rm max}$) appearing 
in equation (\ref{eq:gc_los_int}) is determined by the solid angle observed with $h_{\rm max} 
= r~{\rm{tan}}\phi$ where $\phi$ is the angular resolution of the observation. Performing the 
integration in equation (\ref{eq:gc_los_int}) with an assumed $\sim 10'$ resolution to 
match the Planck data produces the spectrum shown in 
figure (\ref{fig:gc_spec}). As can be seen in that plot nuggets with a baryon number of 
$B\approx 10^{25}$ would saturate the observed haze signal from the inner galaxy. This 
establishes a lower limit on the nugget size based on the Planck data. Note that the 
{\textsc{LPM}} cutoff discussed in section \ref{subsec:spec} does not play a role at the 
frequencies observed by Planck, though can it is shown at low frequencies in  
figure \ref{fig:gc_spec}.

It should be made clear that quark nugget dark matter, while it can reproduce 
the spectrum shown in Fig.\ref{fig:gc_spec}
with $B\sim 10^{25}$ nevertheless cannot explain all the other 
observed features of the haze and, as such, producing the full observed flux 
observed at the galactic centre represents an upper limit of the nugget contribution. 
Thermal emission from the nuggets necessarily tracks the matter density and cannot 
explain the haze emission at large galactic latitudes, the quark nugget spectrum will 
also fail to produce a hard edge to the haze emission as is observed at large latitudes. 
Furthermore the emission from the nuggets will be completely unpolarized, so the 
polarized component observed to trace the edges of the haze emission must be 
produced by other astrophysical mechanisms. Our proposal here is that emission from the 
quark nuggets will provide additional contribution  to the total haze emission at low 
latitudes and, from this picture, to extract limits on the allowed parameter space of 
quark nugget dark matter. 

\subsection{Nearby Milky Way like galaxies}
\label{subsec:Other_galaxies}
\begin{figure}
\includegraphics[width=\linewidth]{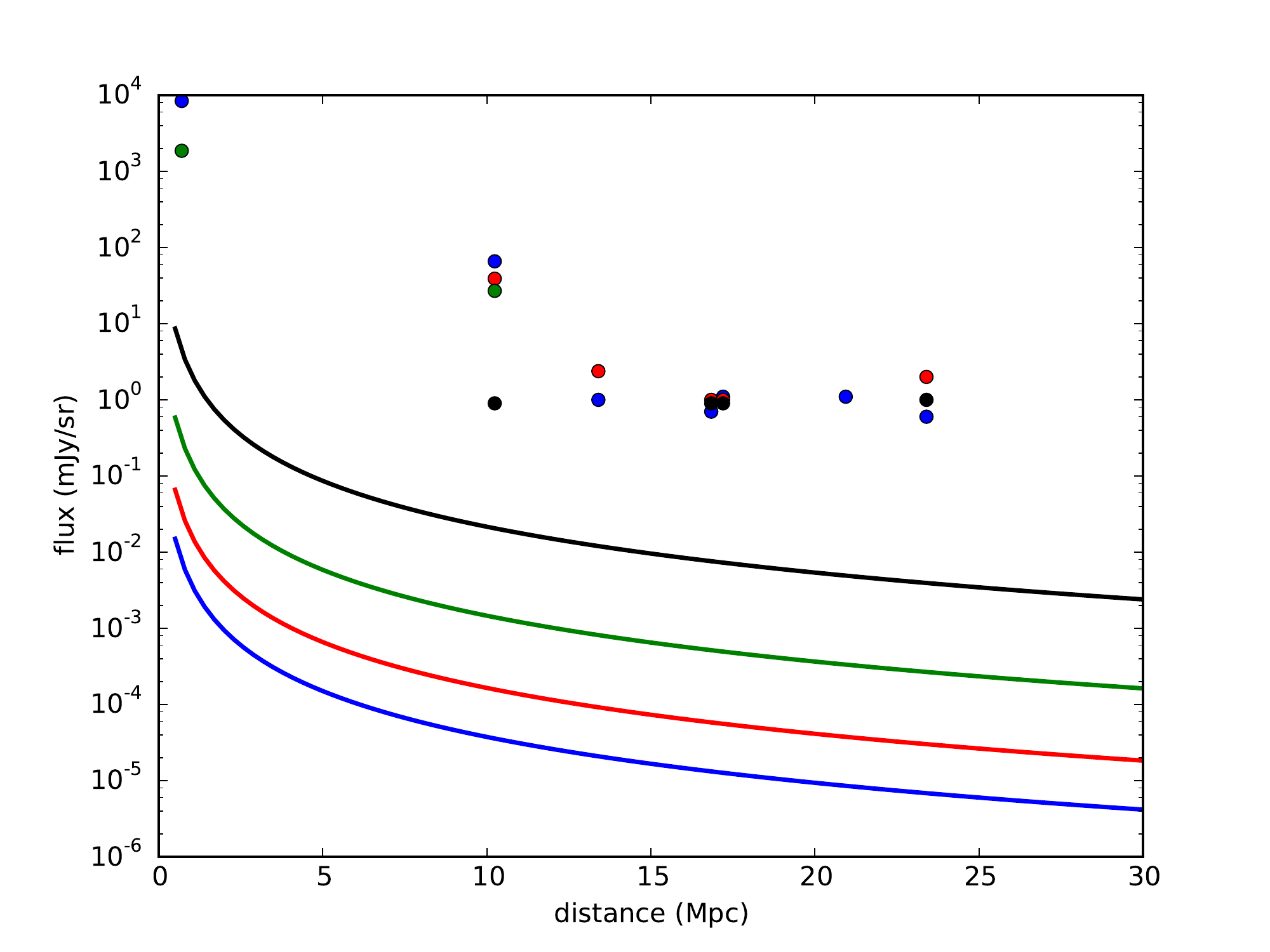}
\caption{Intensity predictions for the quark nugget population of 
a nearby Milky Way like galaxy as a function of distance for a variety of 
wavelengths. Note that the totally emission at a given distance is strongly 
suppresses in the radio bands due to the {\textsc{LPM}} effect as discussed 
in the text. Also shown are a variety of radio observations as used in 
\cite{DM-haze-radio} to constrain a conventional dark matter contribution to the 
galactic haze. Shown are the predicted nugget flux (solid lines) and the observed 
radio signals from nearby galaxies as reported in \cite{DM-haze-radio} (dots) 
the colours indicate frequency with 1.49GHz in blue, 2.38GHz in red, 4.85GHz in 
green and 15GHz in black.}
\label{fig:near_gal}
\end{figure}

Finally we turn to emission from nearby spiral galaxies with matter 
distributions believed to be similar to that of our own galaxy. In this case we 
will determine the total radio band emission from a galaxy. This is done by integrating 
emissivity (equation \ref{eq:nug_emiss}) including the suppression factor (\ref{eq:suppression})  
over the entire matter distribution (this time  
including the disk contribution which may be significant in this 
case as an extended faint disk can 
make a relatively large contribution to total emission.) Once we have established the 
total emission from a spiral galaxy the flux as observed on earth may be obtained from 
the inverse square law. Thus, 
\begin{equation}
\label{eq:gal_flux}
\Phi = \frac{1}{4\pi d^2} 
\int \rm{d}^3r ~\epsilon(r) 
\end{equation}
where, $d$ is the distance to the galaxy. Taking the lower bound obtained from 
the Milky Way observations discussed in section \ref{subsec:MW} we may extrapolate 
the observational consequences for nearby Milky Way like galaxies. For simplicity 
consider a test galaxy with physical parameters identical to those used in our 
discussion of the Milky Way in section \ref{subsec:MW}. We may then translate 
the total intensity in the radio band to a simple distance flux relationship. The results of this 
process, assuming a mean nugget size of $B\sim10^{25}$ which would saturate the galactic 
haze, are shown in Fig.\ref{fig:near_gal}. As can be seen the strong suppression of 
radio band emission results in a galactic radio signal below the observed level in all 
cases. 

This result is in drastic contrast with the studies of \cite{DM-haze-radio} which 
claimed that radio observations of nearby spiral galaxies essentially rule out 
any significant  dark matter contribution to the galactic haze.
The difference of course results from our specific dark matter model in which 
radio emission is strongly suppressed while the emission at CMB frequencies is 
unaffected by the suppression effects studied in sections \ref{subsec:spec} and 
\ref{subsec:others}. The constraints derived in \cite{DM-haze-radio}  , remain fully 
valid for WIMP type dark matter models which are not subject to these 
suppression mechanisms and predict a smooth extrapolation between microwave 
and radio frequencies.   

\section{Conclusions}
\label{sec:con}
It has been demonstrated here that astronomical observation at radio frequencies 
provide only weak constraints on quark nugget dark matter. The fundamental reason 
for this is that thermal emission from the nuggets is suppressed at low energies 
by many body effects within the outer layers of the nuggets. This effect 
is specific to compact composite dark matter models and will not be seen in more 
conventional dark matter models which argue for a haze 
produced by the relativistic products of dark matter annihilations 
or decays. Consequently the strong constraints derived in \cite{DM-haze-radio} are 
entirely valid for WIMP type dark matter models and the suppression effect which 
we discuss here is relevant only in the case of quark nugget dark matter. 

One should note here that  in  previous studies we did discuss  isotropic radio 
emission in the GHz band due to the same quark nugget 
model \cite{Lawson:2012zu}. Furthermore, we claimed in \cite{Lawson:2012zu} that 
the excess in the isotropic radio background at frequencies below the GHz scale  
measured  by  the ARCADE 2 experiment can be naturally explained by the same dark matter 
model studied in the present work. The difference between our present analysis in the radio band 
and our previous study is that the emission 
analyzed in  \cite{Lawson:2012zu}  originated at higher (unsuppressed) frequencies 
but at very earlier times with $z\sim 10^3$ and has subsequently redshifted into the radio, 
this work deals exclusively with the present 
epoch and (strongly suppressed) radio emission originating in the GHz band. 

Across most of the observable parameter space low energy  
suppression comes into effect below the 10-20GHz range. As such the most useful channels 
for investigating quark nugget dark matter are above this scale. For example, improved 
Planck observations of the Andromeda galaxy \cite{Planck-Andromeda} may be able to 
examine a possible haze component from the bulge of Andromeda. Ground based 
radio and microwave observations around 20GHz may also be capable of constraining 
the possible nugget contribution to the spectrum of nearby galaxies, however these 
constraints will be dependent on the exact details of the low energy \textsc{lpm} 
cutoff. 

In conclusion the dark matter proposal advocated in this work may explain a number of 
apparently unrelated puzzles as reviewed in section \ref{subsec:constraints}. All these 
puzzles  independently suggest the presence of some source of excess diffuse 
radiation in different bands ranging over 13 orders of magnitude in frequency. The new element 
highlighted in this paper is that the same DM model is not strongly constrained 
(and certainly, not ruled out) by the analysis \cite{DM-haze-radio}. This is in contrast with vast  
majority of  conventional WIMP's models in which the low energy spectrum continues from 
microwave frequencies to radio frequencies with similar spectral index and whose contribution to 
the haze signal is strongly constrained by \cite{DM-haze-radio}. 

\section*{Acknowledgements} 
We are  tankful to  Ludo Van Waerbeke   who brought our attention to analysis 
\cite{DM-haze-radio}, which eventually initiated  these studies. We are also thankful 
to him for discussions,  questions and comments. 
This research was supported in part by the Natural Sciences and 
Engineering Research Council of Canada.

\end{document}